\documentclass[manuscript]{aastex}

\def\vaa{\hbox{OGLE004336.91-732637.7}}
\def\vbb{\hbox{OGLE004633.76-731204.3}}

\def\va{\hbox{SMC-SC3}}
\def\vb{\hbox{SMC-SC4}}
 \def\gtrsim{\mathrel{\hbox{\rlap{\hbox{\lower4pt\hbox{$\sim$}}}\hbox{$>$}}}}
 \def\ltsim{\mathrel{\hbox{\rlap{\hbox{\lower4pt\hbox{$\sim$}}}\hbox{$<$}}}}

\newcommand{\myemail}{rmennick@astro-udec.cl}

\slugcomment{Submitted to PASP.}

\shorttitle{Photometry of two unusual A supergiant systems in the SMC}
\shortauthors{Mennickent et al.}

\begin{document}


\title{Photometry of two unusual A supergiant systems in the Small Magellanic Cloud}

\author{R. E. Mennickent\altaffilmark{1}}
\affil{Departamento de Astronom\'{\i}a, Facultad de Ciencias F\'{\i}sicas y
Matem\'aticas, Universidad de Concepci\'on, Casilla 160-C, Concepci\'on, Chile}
\email{\myemail}

\author{M. A. Smith}
\affil{ Department of Physics, Catholic University of America, Washington, DC 20064, USA; Present
address: Space Telescope Science Institute, 3700\\ San Martin Dr., Baltimore, MD 21218, USA }

\author{Z. Ko{\l}aczkowski}
\affil{Instytut Astronomiczny Uniwersytetu Wroclawskiego, Kopernika 11, 51-622 Wroclaw, Poland}

\author{G. Pietrzy{\'n}ski}
\affil{ Warsaw University Observatory, Al. Ujazdowskie 4, 00-478 Warszawa, Poland}

\and 

\author{I. Soszy{\'n}ski}
\affil{ Warsaw University Observatory, Al. Ujazdowskie 4, 00-478 Warszawa, Poland}

\begin{abstract}
We present multiwavelength broadband photometry and $V, I$ time resolved photometry for two variable bright stars in the SMC, OGLE004336.91-732637.7 (\va) and \vbb~ (\vb). The light curves span 12 years and  show long-term periodicities (SMC-SC3) and modulated eclipses (SMC-SC4) that are discussed in terms of wide-orbit intermediate mass interacting binaries and associated  envelopes. 
SMC-SC3 shows a primary period of 238.1 days  along with a complicated
waveform suggesting ellipsoidal variablity influenced by an eccentric orbit.
This star also shows a secondary variability with an unstable periodicity 
that has a mean value of 15.3 days. 
We suggest this could be associated with nonradial pulsations.

\end{abstract}


\section{Introduction}

The evolution of massive stars is an important topic of stellar 
astrophysics because it provides clues to the mechanisms that feed the
Galactic medium with the building blocks of future generations of stars. 
Many of the evolutionary stages of massive stars are short-lived and hence 
challenge our ability to find enough examples of a common group to
characterize them.
Detecting objects in these brief phases of evolution is of great use in
testing current theories of massive star evolution, especially if they
represent populations with low metal abundances, a property of stars  
in the
Magellanic Clouds.
A complete census of pre- and just post-main sequence population of massive 
stars in the Clouds has not yet been compiled, and thus attributes of these
subpopulations are not well known, but they are known to be generally variable.
Catalogs of B stars with unusual variable light curves in the 
Magellanic Clouds (e.g., Mennickent et al. 
2002, hereafter M02 and Sabogal et al. 2005), based on OGLE photometry (Udalski et al. 1997, Szyma\'nski 2005),
provide
excellent material for analysis of massive stars near the main sequence.  
In an effort to  characterize some of these stars, follow
up spectroscopy was conducted for two stars in the Type\,3
M02 sample\footnote{Type-3 variables are SMC Be star candidates with $I$-band light curves varying periodically or quasi-periodically.}, that will be given in a future paper (Mennickent \& Smith 2010, hereafter MS10).  Both these stars turn out to be probable binaries and to
have highly peculiar characteristics. In this paper we present the
photometric results for these stars.

 The stars chosen were taken from an initial  sample of 8 Type-3 variables in
the M02 catalog satisfying the arbitrary criterion of having visual magnitudes 
brighter than 14.2. 
No other criteria were imposed on their selection.
The stars selected were \vaa~  ($\equiv$ SMC-SC3-63371, MACHO ID 213.15560; 
hereafter \va) and \vbb~ ( $\equiv$ SMC-SC4-67145, MACHO ID 212.15735.6; 
hereafter \vb).  Exploratory  optical spectra  exhibited 
substantial 
H$_{\alpha}$ emissions in both objects, and in the case of SMC-SC3, multiperiodic variability (Mennickent et al. 2006, hereafter M06). 
 \va~ was recently included in the slitless survey of H$\alpha$ emission line objects by Martayan et al. (2010, their star in cluster SMC-17). 


The goals of this paper are to characterize the photometric properties 
of these two stars with the longer time baseline available and gain further 
insights on the formation and nature of these systems.

\clearpage

\begin{deluxetable}{cccccccccc}
\tabletypesize{\scriptsize}
\tablewidth{0pt}
\tablecaption{$UBVR$ magnitudes from Massey (2002) and OGLE photometry 
are given. f Dereddened $B-V$ colors and derived spectral types are also included.
}
\tablehead{
\colhead{Object} & \colhead{U} & \colhead{B} & \colhead{V} & \colhead{R}&  \colhead{V$_{OGLE}$} & \colhead{(B-V)$_{OGLE}$} & \colhead{(V-I)$_{OGLE}$}  & \colhead{(B-V)$_{0}$} &Sp. type }
\startdata
\va~ & -&  13.63&  13.48&  13.30& 14.18&  0.181& 0.331& 0.08  &A4\\
\vb~ & 14.34& 14.15&  13.94 & 13.70 & 14.06&  0.206& 0.385& 0.11&A5  \\
\enddata
\end{deluxetable}

\clearpage

\clearpage

\begin{deluxetable}{cccccccc}
\tabletypesize{\scriptsize}
\tablewidth{0pt}
\tablecaption{Infrared magnitudes for program stars. 
Phases refers to ephemeris given in Equations 1 and 4.}
\tablehead{
\colhead{Object} & \colhead{I} & \colhead{J} & \colhead{H} & \colhead{K}&  \colhead{JD/Date} & \colhead{Phase} & \colhead{Source}  }
\startdata
\va~  &13.701(9) &13.346(22) &             -&12.954(116) &1998-08-12  &0.52   & cds.u-strasbg.fr/denis.html\\ 
\va~  &13.847(30)&13.472(90) &             -&13.050(180) &2450414.6148&0.90 & cds.u-strasbg.fr/denis.html \\
\va~  &13.781(40)&13.560(110)&             -&13.134(160) &2451048.7763&0.56   & cds.u-strasbg.fr/denis.html \\
\va~  &-         &13.545(42) &13.341(50)    &13.275(40)  &2451034.7109&0.50   & www.ipac.caltech.edu/2mass/\\ 
\va~  &-   &13.540(20)& 13.380(10)& 13.180(20)&-&-&pasj.asj.or.jp/v59/n3/590315/\\
\vb~  &13.655(30)&13.388(90) &             -&13.316(210) &2450418.5524&0.42& cds.u-strasbg.fr/denis.html \\ 
\vb~  &13.616(30)&13.383(130)&             -&13.079(150) &2451039.7991&0.79& cds.u-strasbg.fr/denis.html \\
\vb~  &-         &13.403(29)&13.236(34)     &13.054(35)  &2451034.7134&0.76& www.ipac.caltech.edu/2mass/\\
\vb~  &-   &13.380(10)& 13.260(10)& 13.170(20)&-&-&pasj.asj.or.jp/v59/n3/590315/\\
\enddata
\end{deluxetable}

\clearpage

\section{Photometric results}
\label{pht}

 We give broad-band $V$ and $V_{OGLE}$ magnitudes and $UBVR$ colors
for both program stars  in Table 1, along with dereddened colors $(B-V)_{0}$ and estimated spectral types. 
 For \va,  we calculated $(B-V)_{0}$ using $E(B-V)$= 0.10 (Martayan et al. 2010).
For \vb,  we assumed $E(B-V)$= 0.09 as representative for SMC (van den Bergh 2000). The spectral types listed in Table\,1 were derived from $(B-V)_{0}$ using the 
Fitzgerald (1970) calibration and assuming a luminosity class of II or I, derived from their magnitudes and membership in the SMC. The $(B-V)_{0}$ color 
suggests middle A spectral types. The previous spectroscopic 
classification of A7-F5\,e for SMC-SC3 and F5\,Ie+G0\,I for SMC-SC4 given by M06 was probably  influenced by the 
 detection of shell metallic lines in low resolution spectra, that are formed in relatively cool stellar envelopes.

 From the visual magnitudes and SMC distance (Udalski 2000), we can
estimate the absolute magnitude as 
$M_{V}$ $=$ -4.83  for SMC-SC3 and -5.05 for SMC-SC4. 
 Note that this refers to the visual magnitude of the primary since
the secondary contributes only a few percent to the light in this region.
Using a bolometric correction of  -0.13 for a A5 supergiant, we obtain 
 $M_{bol}$ = -5.0 and -5.2. 
From the evolutionary tracks for stars with SMC metallicities by Meynet \& 
Maeder (2001, hereafter MM01),  both stars fit tracks of 9 M$_{\odot}$ 
evolved stars in the MM01 models.  
As the evolutionary tracks of MM01 depend on the value of stellar mass and rotation velocity, 
the position of the optical component of these systems in the $M_{V}-T_{eff}$ diagram cannot be used with their models
to alone determine the evolutionary state.

\begin{figure*}
\centering
\scalebox{1}[1]{\includegraphics[angle=0,width=15cm]{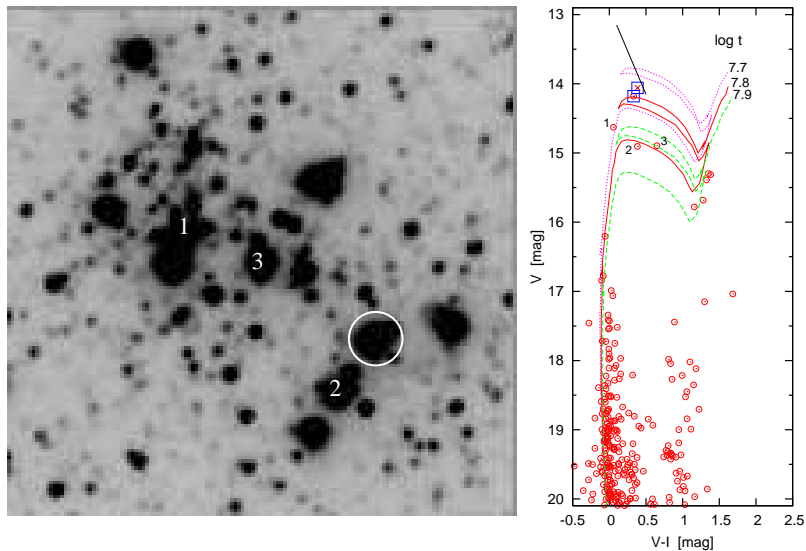}}
\caption{The left image is a finding chart of \va\ centered on the 
cluster NGC\,242 (1' $\times$ 1' subframe of the reference $I$-band 
image from the OGLE-II survey).   North is up and East left. The position of 
\va~  is given by a circle  and three reference stars are labeled with numbers.  Right panel 
shows the CMD for the same region. The CMD is based on the OGLE\,II standard 
$VI$ photometry in the SMC (Udalski et al. 1998a). Three isochrones are also 
presented (Bertelli et al. 1994) for stars with metallicity $Z$= 0.004
and ages ($\log t$) labeled in the picture. 
We adopted these isochrones using $V-M_{v}$= 19.0, $E(V-I)$= 0.08 and
$A_{v}= 2.5*E(V-I)$.  The reddening vector is indicated by the upper 
oblique line. For comparison we indicate SMC-S04 with a cross in an open square   at the upper right of the SMC-SC3 position.}
  \label{1}
\end{figure*}

SMC-SC3 is a probable member of the open cluster NGC\,242, and in fact 
its  angular distance from the cluster center is only $\sim$ 10". 
In Figure\,1 we show the color-magnitude diagram for \va~ and for 
neighboring stars present in their field. We investigated the light 
curves for the stars labelled 1, 2 and 3, and we found them to be
photometrically constant. The figure also depicts isochrones for the 
cluster NGC\,242 considering the parameters given in the figure caption.  
It is clear that 
 the bright objects represented here lie close to the $\log\,t$ 
=  7.8 isochrone.  This is approximately the reported age of this cluster 
(7.9; Ahumada et al.\,2002). 
Fig.\,1 also depicts the observed position of \va.~ Its unreddened
location in the HR Diagram should be projected back along the indicated 
reddening vector.  The position of both objects in the HR Diagram indicates that they are
evolved luminous stars found in a stage during core hydrogen exhaustion.

  The 2MASS $J-K$ photometry of these two stars is minimally affected by interstellar reddening.
These values are 0.27 and 0.35 for SMC-SC3 and SMC-SC4, respectively (Table 2), much larger than the typical $J-K$ color of an A5 supergiant (0.12 mag, Koornneef 1983), as inferred from the $(B-V)_{0}$.  Similarly
the $V-K$ colors, of about  0.5 and 0.8 mag, are much larger than the typical $V-K$ color of an A5 supergiant (0.36 mag, Koornneef 1983). 
Emission caused by scattering of free electrons in circumstellar
envelopes is a likely candidate for explaining the infrared excess.  Their moderate value excludes the presence of much dust around these stars
and explains the discrepancy between spectral types derived from optical and infrared photometric colors.

Deep exposures of sky images surrounding the two program objects 
reveal no suggestion of nebulosity. However,
during our investigation we discovered that \va~ has a visual
companion, the small bump 1" NW of \va~ in Fig.\,1.
We analyzed its OGLE light curve as part of this study and 
discovered that this visual companion is an
eclipsing binary with an orbital period of 2.96934 days, $V$= 16.776, and
$V-I$= -0.083. Its $\Delta I$ range of variability range was 0.77 mag. 
This nearby star is not in the catalog of SMC eclipsing binaries  
(Bayne et al. 2002) but was included in the catalog of eclipsing variables by Udalski et
al. (1998b) under designation SMC-SC3 star number 63551.
 The fact that the star is 2.6 mag fainter than SMC-SC3 makes it unlikely that the
   periodicities discussed below
   are caused by the contribution of 
   this star's light to the photometric colors.

 We have included the following datasets in our analysis: 
 OGLE-II ''DoPhot" $V$ light curves, OGLE-II ''DIA" $I$ light curves and OGLE-III "DIA" $V, I$ light curves. A summary of these datasets is given in Table 3.
The range of HJDs shown corresponds roughly from mid 1997 to mid 2009.

We proceed to analyze the photometric variations in 
our time series, using the {\it pdm} (after the Phase Dispersion
Minimization algorithm; Stellingwerf 1978).  

\subsection{Characterization of the light curve of \va}

The expanded dataset 
allowed us to examine the possible range of periods 
over a greater range than was available to M02 and M06.  In fact, we found
a more reliable primary period that is twice as long as 
these authors had found, namely 238.1 (+2.3,-3.1) days,
but it clearly gives the deeper minimum in the PDM periodogram.
The new baseline in time is sufficient to
rule out a yet longer (e.g., 476 day) period.
The period error is the $HWHM$ of the (asymmetrical) periodogram's peak.  
The improved ephemeris for the measured centroid of the light curve 
maximum of \va~  is:\\ 

$T (HJD) =  2\,450\,915.3 + 238.1 (+2.3,-3.1)\,E.$\hfill(1) \\

\noindent 
We modeled the \va~ light curve with harmonics and subharmonics of this fundamental period and, after analyzing residuals, it was clear that,   apart from a small seasonal  variability, the second periodicity of 15-days, reported by M06,  still persisted  in the Fourier spectrogram (Fig.\,3). For this  sinusoidal 15 day periodicity we found the ephemeris:\\

$T (HJD) =  2\,452\,739.76 + 15.35(0.02)\,E.$\hfill(2) \\

\clearpage
\begin{deluxetable}{lrrrr}
\tablewidth{0pt}
\tablecaption{Summary of photometric data analyzed in this paper. HJD zero point is
 2400000. OGLE 1, 2 \& 3 data are considered.}
\tablehead{
 \colhead{Object} & \colhead{Dataset} & \colhead{HJD-start} & \colhead{HJD-end} & \colhead{N-obs} }
\startdata
SMC-SC3 &OGLE I-band &50621.83606 &54866.55206  &1022\\
SMC-SC3 &OGLE V-band &50670.89064 &54792.59219  &82\\
SMC-SC4 &OGLE I-band &50621.79700  &54954.88836 &1067\\
SMC-SC4 &OGLE V-band&50645.91866&54954.89402&94\\
\enddata
\end{deluxetable}

\clearpage

\noindent
 
 The Fourier periodogram of \va~ indicates
that sidelobes surround the primary peak. This fact suggests the possibility that the Fourier spectrum shows the combined effect of several  close frequencies acting simultaneously giving origin to harmonic interactions. However, in this case we should observe a modulation of the amplitude of the light curve, and this is not observed 
 in our reconstruction of the 15-day light curve.
Another interpretation for the existence of sidelobes comes  from the analysis of the
 O-C diagram (based on observed minus calculated 
ephemeris, e.g. Sterken 2005),  that can be used  as a diagnostic for a constant 15-days signal
(Fig.\,4.) The O-C diagram should display 
the differences between predicted and observed phases from cycle to cycle 
as following a horizontal or sloped line if the variable had a constant 
period. On the contrary, most of the data points in  Fig.\,4  follow a 
 regular oscillation for the  15-day period  - that is they suggest a single period that varies over time.
 The period changes around a mean value  on a timescale  of $\approx$ 3800 days.
 
  In order to test the possibility  that the sidelobes observed in the Fourier spectrum of the residual light curve  indeed arise from a slowly changing period we constructed synthetic light curves represented by a slow oscillation 
around 15 days with "superperiod" of 3800 days,\,viz. :\\

$ I = 13.84 + A \sin(\frac{2 \pi t}{15.32 + B \sin(2 \pi t/3800)})~,$\hfill(3)\\  

\noindent
where $t$ is the observing time in days and $A$, $B$ constants to be adjusted to the observations,
typically  0.02 mag and 0.15 days, respectively.

  The corresponding Fourier spectra showed two sidelobes around the main frequency, suggesting in fact that the Fourier spectrum of residual light in SMC-SC3 is consistent with a variable period. The general appearance of the observed O-C diagram was also reproduced by our simulation.  However, we emphasis
that this representation, including the persistence of a 3800-day superperiod,
has no predictive power for new additions to this star's light curve.
   
  In the following, and only as a matter of convenience for the analysis, we will consider this complex variability as the simple periodicity represented by the ephemeris given in Eq.\, 2.

  Whereas the 15-day cycle is represented  roughly by sinusoidal variability, the 238 days periodicity is characterized by two unequal minima, like those observed in ellipsoidal variables (Fig.\,5).  However SMC-SC3 does not appears to be a bonafide ellipsoidal variable. The maximum of its light curve is a sharp excursion over a longer asymmetrical variation and the first rising branch is steeper and shorter than the second one. All these features, especially the double modulation and the two unequal minima not separated by  half a cycle, are consistent with ellipsoidal variations in an atypical {\it eccentric-binary} (Hilditch 2001).

\subsection{Characterization of the light curve of \vb}

  The light curve of SMC-SC4 is of interest in that it not only
produces eclipses but the eclipse depths can vary from cycle to cycle
 (Fig.\,2).  The ephemeris for the eclipses is: \\

$T_{min} (HJD) =  2\,450\,709.9(2) + 184.26(1.25)\,E, $\hfill(4) \\

\begin{figure*}
\centering
\scalebox{1}[1]{\includegraphics[angle=0,width=10cm]{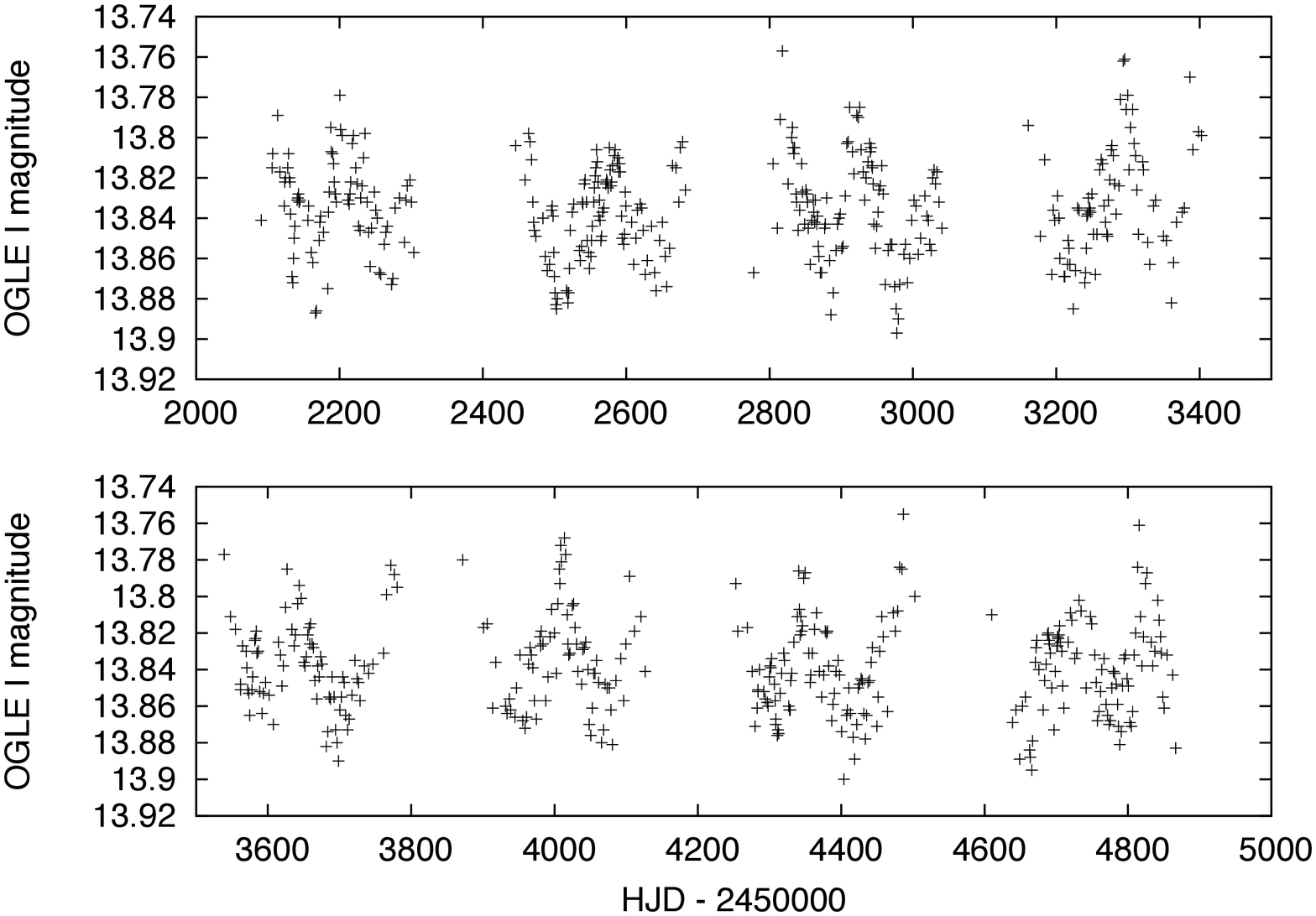}}\\
\scalebox{1}[1]{\includegraphics[angle=0,width=10cm]{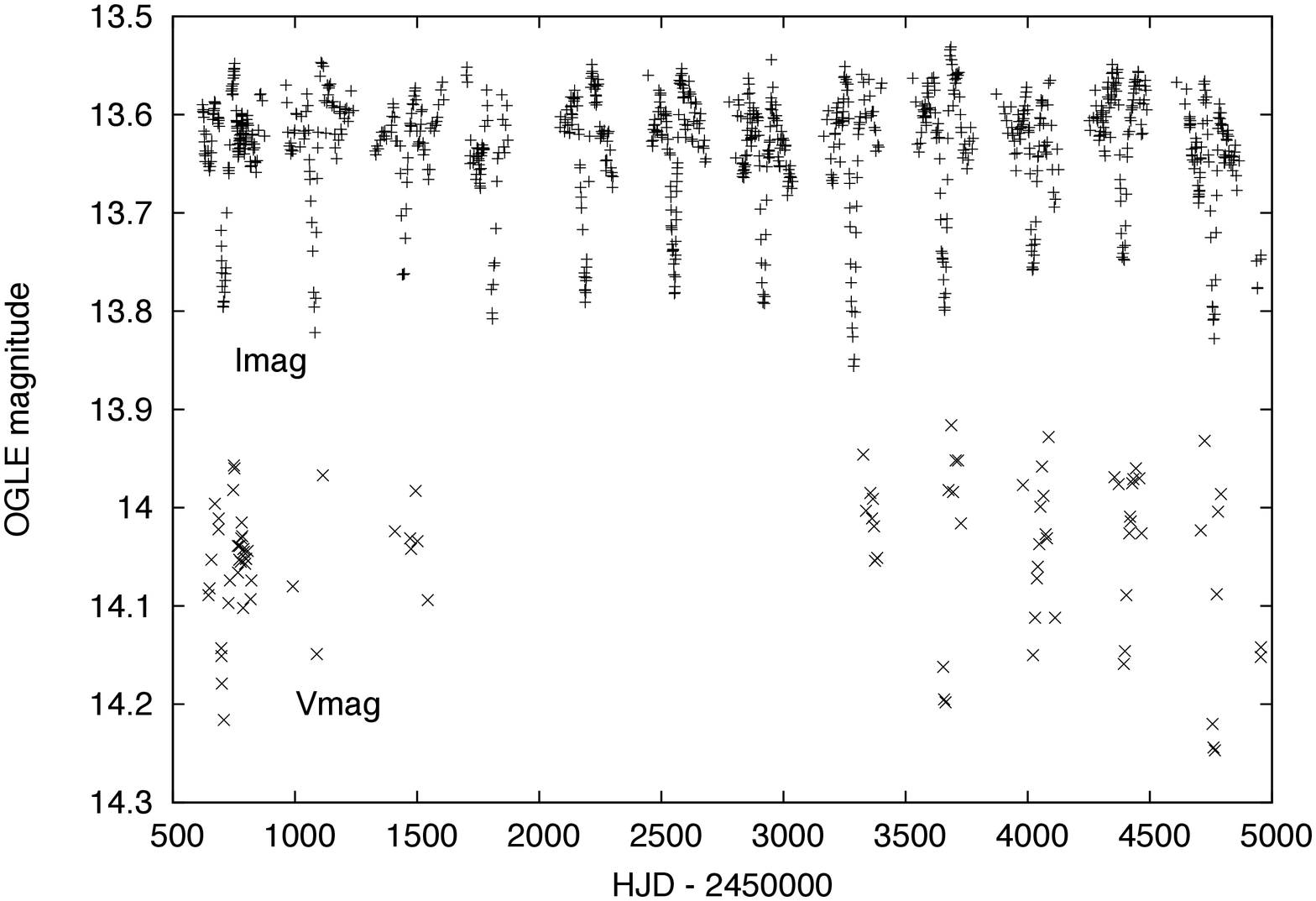}}\\
\caption{OGLE  I-band light curve of SMC-SC3 (up) and SMC-SC4 (below). Typical formal  error is 0.004 mag.}
  \label{2}
\end{figure*}

\begin{figure*}
\centering
\scalebox{1}[1]{\includegraphics[angle=0,width=10cm]{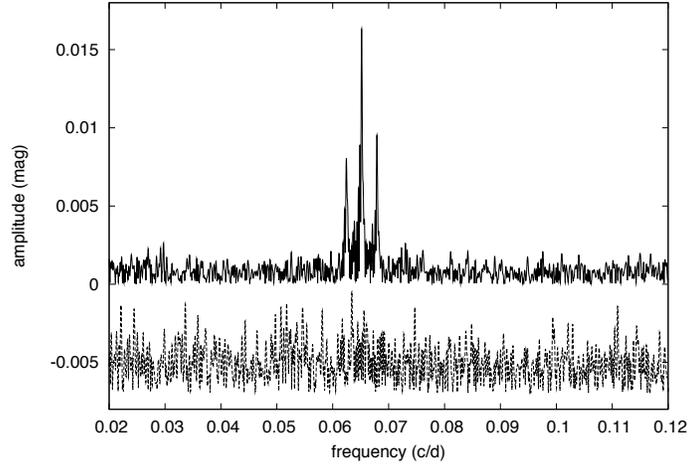}}\\
\caption{Fourier spectrum of such a residual light curve along with the scaled and vertically shifted window spectrum. The main peak is the 15-days periodicity.}
  \label{3}
\end{figure*}

\begin{figure*}
\centering
\scalebox{1}[1]{\includegraphics[angle=0,width=10cm]{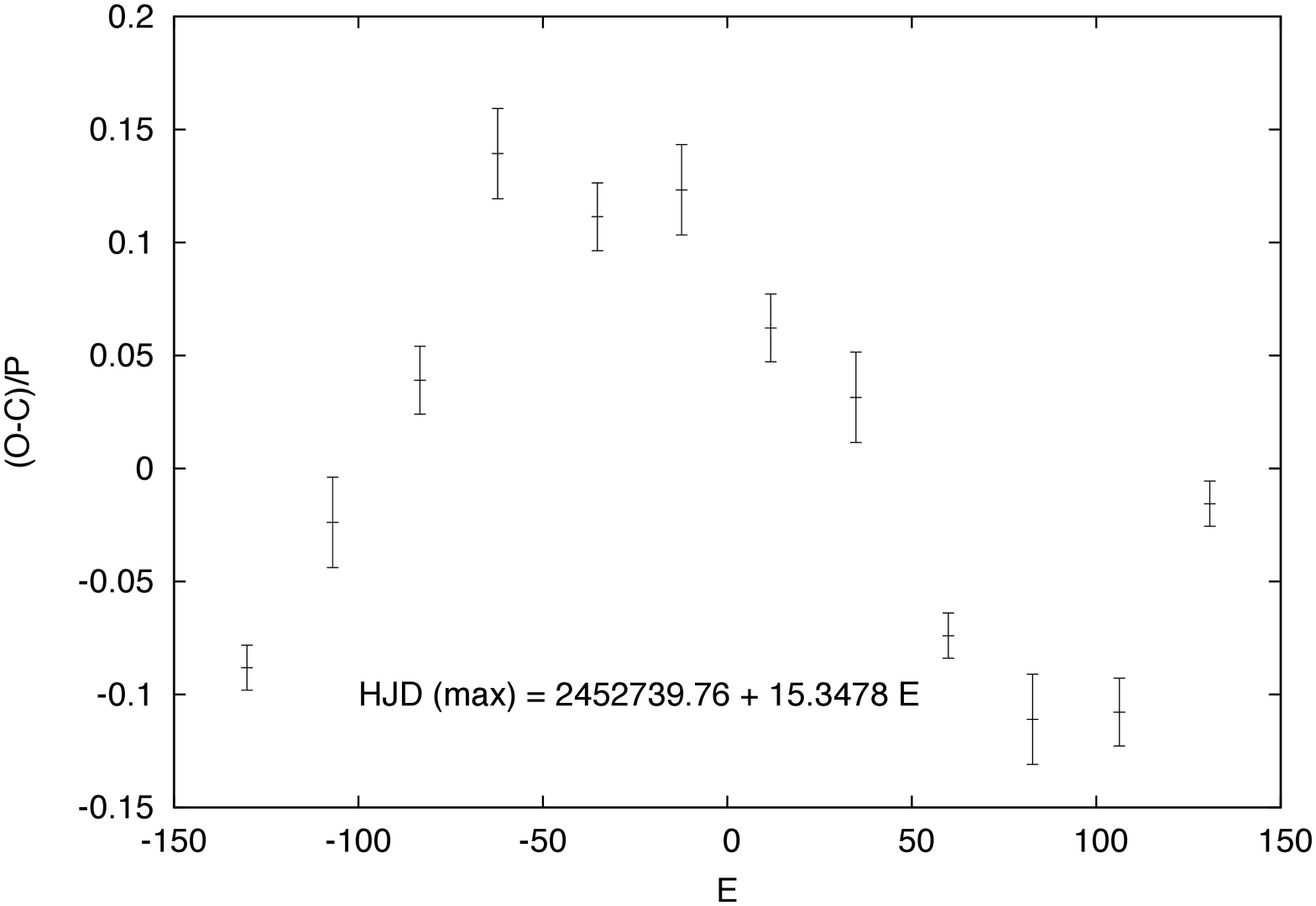}}\\
\caption{ O-C diagram for the maxima of the 15-d periodicity of \va.
This figure shows a time scale of variability for the 15-days cycle of 3800 days, corresponding to twice the distance between maximum and minimum (about 250 cycles).  }
  \label{4}
\end{figure*}

\noindent
 where we used  the {\it pdm} algorithm on the expanded I-band  
dataset to search for periods.
 From the median of the absolute values of the differences between magnitudes of 
point to point
measurements we estimate a characteristic noise of 0.01 mag for the SMC-SC4 light curve. 
However, the variability actually observed in the light 
curve indicates that the variability discussed here is a signal from the star.
The new analysis also shows that the former 24-day periodicity reported by 
M06 was  the result of their time-limited dataset
and the presence in the light curve of non-coherent quasiperiodic
modulations, perhaps pulsations. This signal is responsible for the 
large noise visible in the smoothed curves of Fig.\,6a, particularly 
outside the primary eclipse. This signal  also 
illustrates the absence of a secondary eclipse in the $I$-band.
The $V$-band, however, suggests the presence of a very wide secondary eclipse
(Fig.\,6b). Furthermore, we conclude from this figure that the eclipses
are irregular in shape and depth, although they recur with a {\it mean}
period of 184 days. They are about as deep in $V$ as in $I$,
 0.5-0.6 magnitudes. Changes in the eclipse  shapes and depths occur
and these  last appear to be modulated.
For instance, OGLE III data suggest a supercycle  for the eclipse depth of about eight times 
the basic period, but this tendency disappears when considering the
earlier OGLE II dataset (i.e., those obtained before HJD\,2452000, Fig.\,2).
We argue  that the same low frequency variability observed outside eclipse 
could be responsible for the observed changes in eclipse shape, but not for the depth changes.

\section{Discussion}

In this section we will evaluate the interesting attributes of SMC-SC4 
light curve first, and then follow with a discussion on the even more 
remarkable properties of SMC-SC3.

The fact that eclipses of SMC-SC4 are irregular in depth 
but with an arguably regular timing indeed
indicates that the A star is eclipsed by an almost opaque but possibly 
ever changing body, rather than by the secondary star. 
There are at least two additional reasons that the eclipse cannot reasonably 
be ascribed to a star-star eclipse:

\noindent i) a large fraction of the A star is eclipsed, and if they
were caused by the much hotter star  (such as we observe in the 
near-UV, see MS10) the eclipse depths would be 
larger in the $V$ magnitude than in $I,$

\noindent ii) the radius of the secondary star is likely to be smaller 
than the A star's and cannot account for the large geometric obscuration
observed in the $I$ band,
yet the long ($\sim$0.2 cycles) duration of the eclipse implies that 
a large body orbits near the A star. 
This body is  warm (accounting for the nearly equal eclipse depths 
in the two photometric bands) and is therefore likely 
to be within a stellar radius or so from the A star.

We conclude that that the eclipses are inconsistent with a
third star in a dynamically stable orbit.  Rather they are
caused by a large  almost opaque body probably extending well out of the 
orbital plane of the SMC-SC4 system  and intruding to our line of sight 
during this phase.
It appears to be due to an impermanent, ever-replenished structure 
that co-orbits within the  binary star system.  

The photometric variability of SMC-SC3 analyzed in this paper consisting of a main ellipsoidal variability is consistent with a binary nature. As mentioned earlier, the double minima of the 238-day folded light curve exhibit unusual signatures which can be interpreted only in
terms of an ellipsoidal variable. The unequal amplitudes, spacings, and
asymmetries of the minimum lobes strongly suggest that this 
object  is a binary with a mild eccentricity.
Model light curve grids of  Soszy{\'n}ski et al. (2004) 
suggest that the asymmetry of especially the  time of second minimum is 
sensitive to a large orbital eccentricity. A comparison of these models imply
that the orbital eccentricity is nonzero but modest, e.g. $e$ $\approx$ 0.2.
However, another diagnostic argues that any eccentricity in this system
is actually somewhat larger. In particular,
the sharp maximum observed for SMC-SC3 at phase 0.0 (Fig.\,5) 
is not typical of small eccentricities,  but examples have been
well documented. For example, a
similar feature has been observed, but much less pronounced,  
in V380\,Cygni (Guinan et al. 2000, see other examples in Fig.\,8 
of Soszy{\'n}ski et al. 2004). These maxima 
are
frequently interpreted as the reflection of the more hotter
star on the distorted and more distended cooler star at minimum 
binary separation.
According to canonical interpretations of ellipsoidal variable light
curves, the light maximum corresponds to a time when the visible star
(in this case, the secondary at optical wavelengths) presents its maximum
area in the plane of the sky. The fact that we see evidence in the form
of an absolute light maximum at this same phase indicates that this
phase corresponds nearly to periastron passage as well.
Although this places some geometrical constraints on the orbital
parameters,  it is clear 
that a precise determination of the orbital parameters requires 
a set of RVs sampling the whole binary cycle. Most importantly, given
a period as long as 238 days, the semimajor axis of the orbit must be
large, and this, together with the presence of a reflection effect
suggests that the eccentricity is large, perhaps $e$ $\approx$ 0.9.

\begin{figure*}
\centering
\scalebox{1}[1]{\includegraphics[angle=0,width=10cm]{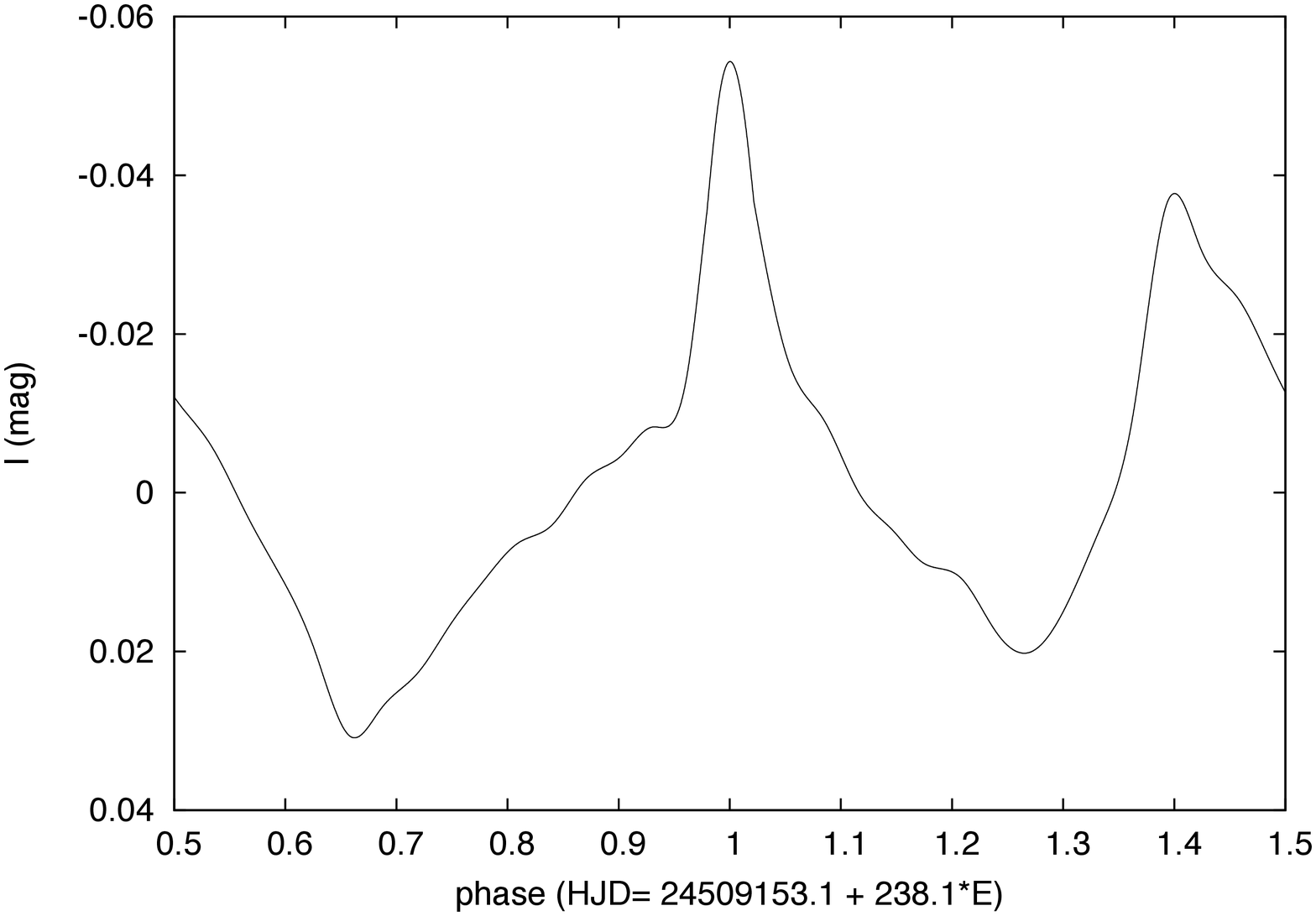}}\\
\caption{The spline function representing the ellipsoidal variability of SMC-SC3  folded over a  period of 238.1 days.   Spectra to be discussed
in M10 are at phases 0.28, 0.69 and 0.44.}
  \label{5}
\end{figure*}

\begin{figure*}
\centering
\scalebox{1}[1]{\includegraphics[angle=0,width=10cm]{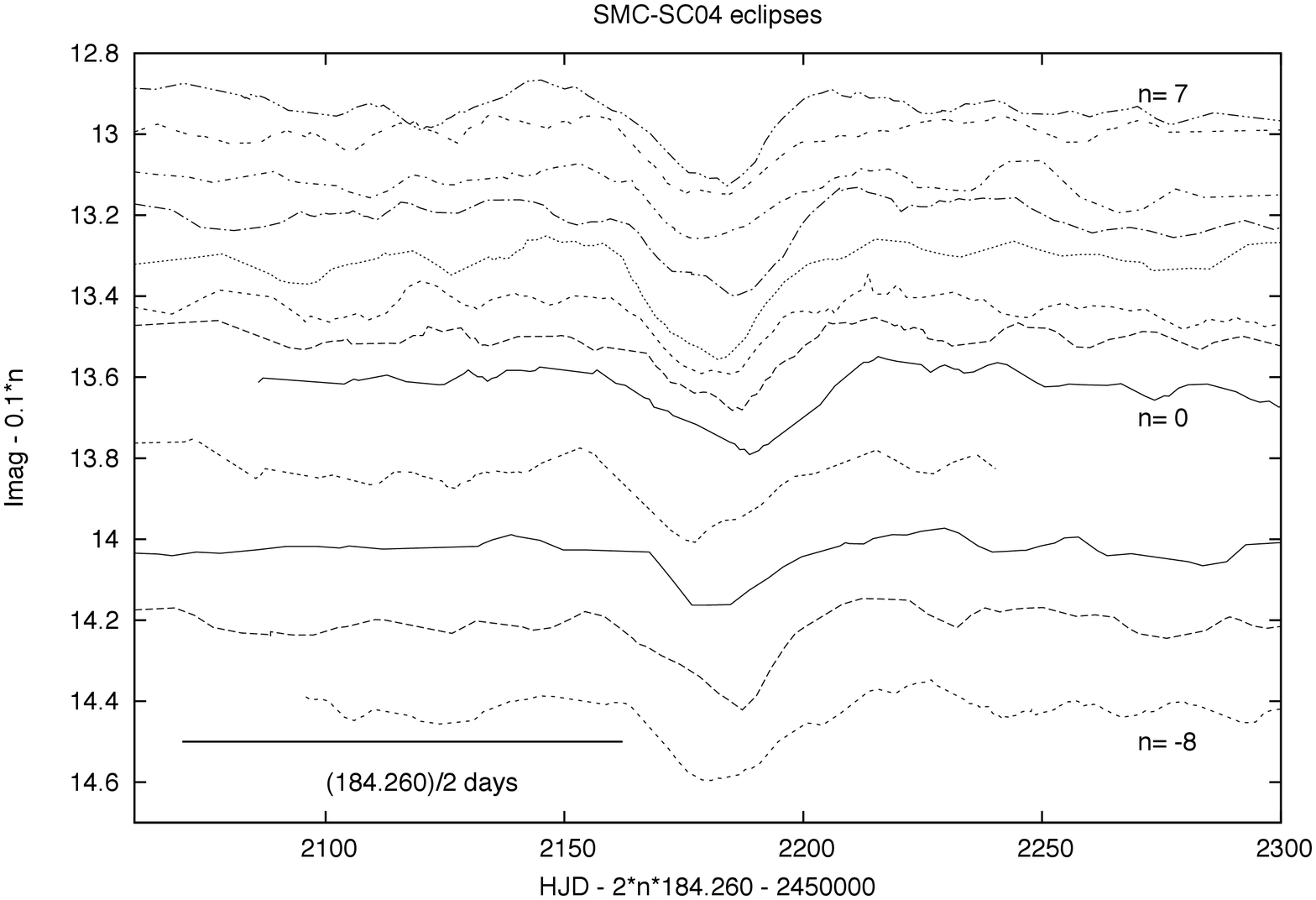}}\\
\scalebox{1}[1]{\includegraphics[angle=0,width=10cm]{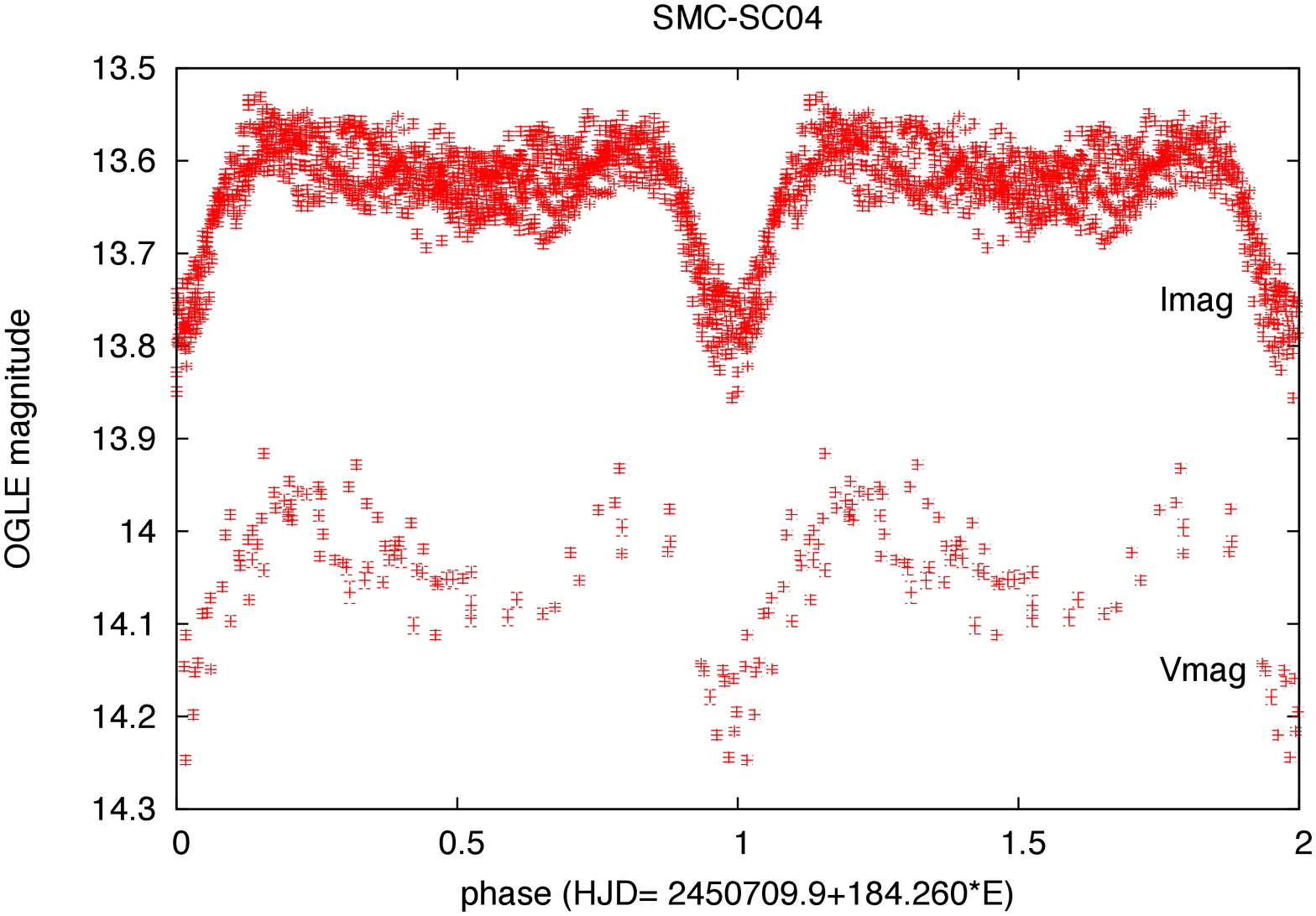}}\\
\caption{ Coplotting of eclipses in SMC-SC4 (upper panel). The factor ``2" in the x-axis label accounts for the data 
seasonal gaps.  The star was observed annually at the same time of year, and one season happens to be roughly twice the orbital period of the star. 
Light curves for $I$ and $V$ filter OGLE data 
are folded with the ephemerides given in equation 2 of the text, i.e. with the period of 184.25 days (bottom 
panel).  The $V$ band data were taken over part of the longer timespan of the $I$ band observations.
Note that the eclipse minima in $V$  are at least as deep as for the $I$ filter.  Spectra to be discussed
in M10 are at phases 0.10, 0.23 and 0.66.}
  \label{6}
\end{figure*}

 We are not yet able to reconcile these considerations, but we have
a preference currently for the high eccentricity because we believe
the "reflection effect" is the more robust interpretation.

We note the similarity of the photometric behavior of our targets with 
the recently released light curves of OGLE-LMC-LPV-41682, that with $V$= 13.958
shows eclipses of variable depth with period 219.9 days, and a second periodicity of 44 days, and possibly OGLE-LMC-LPV-15046, that with $V$= 16.887  shows a main period of 
148.67 days, along with lower amplitude periods of 272.60 and 181.88 days, resulting in a light curve with oscillations of variable amplitude  
(Soszy{\'n}ski et al. 2009). 
This opens the possibility that SMC-SC3 and SMC-SC4 are not 
rarities and can be understood within the context of luminous 
interactive binaries in which mass loss and/or exchange may be occurring.
These objects  will be discussed in a forthcoming paper.

Regarding the short photometric cycles observed in both systems, 
(15-days cycle in SMC-SC3 and  otherwise still unspecified 
non-coherent variations in SMC-SC4) 
we  consider the possibility that they could be linked to nonradial pulsations of the A 
supergiant.  The variability of the 15-day photometric cycle observed in SMC-SC3  suggests a 
non-binary nature. 
 It is generally believed that hot 
and luminous stars of the $\alpha$-Cygni type show irregular 
variability driven by nonradial pulsations on time scales of weeks  
(e.g., de Jager et al. 1991).
These are supergiants of spectral types B-A and amplitude of 
variability $\sim$ 0.1 mag. It is then possible that part of the 
photometric variability observed in our program  objects corresponds 
to pulsational activity. The amplitude of these pulsations in \vb~ 
is large enough that we suspect that even partial 
eclipses of the A-type supergiant allows their detection.

We considered also the possibility that the 15-day cycle corresponds 
to advection of stellar spots.  However, for  $v\,sin\,i$= 20 $\pm$ 5 km/s and $R$= 
30 $R_{\sun}$ derived by MS10,  the rotational period of the A 
supergiant should be 76\,sin\,i days. This is  too long to fit the 15-day
periodicity
and, assuming that its orientation is
not nearly pole-on, not synchronous with the binary period.

\section{Conclusions}

We have presented the analysis of OGLE light curves of  two SMC 
bright systems showing novel photometric 
properties. These properties are consistent with the interpretation that these stars are long-period interacting binaries with an evolved most-luminous stellar component. We find 
eclipses in SMC-SC4
with P$_{o}$= 184 days modulated in depth and  perhaps  shape on time scales of hundreds of days, suggesting the presence of a variable and non-stellar eclipsing region.
In addition, we discovered an unusually strong reflection effect  in the orbital light curve  of SMC-SC3 (P$_{o}$= 238 days) and a short variability with quasi-period 15.3 days changing on time scales of 3800 days. We note the possibility that the short-term fluctuations observed in both stars 
are  signatures of  nonradial pulsation.  This may explain the ``drifting" of a single 15-day periodicity in the 
light curve of SMC-SC3.

\acknowledgments

 We thank an anonymous referee for a rapid review and useful indications that improved an earlier version of this paper. 
REM acknowledges support by Fondecyt grant 1070705, the Chilean 
Center for Astrophysics FONDAP 15010003 and  from the BASAL
Centro de Astrof\'isica y Tecnologias Afines (CATA) PFB--06/2007.
We thank Dr. Darek Graczyk for discussions about the reflection effect observed in SMC-SC3.
\\

\clearpage

\end{document}